\documentclass{article} 

\usepackage{graphicx}
 \usepackage{epstopdf}
  \usepackage{color}
	\usepackage{amsmath}
	\usepackage{amsthm}
	\usepackage{amssymb}
\usepackage{enumitem}
  
\setlist[enumerate]{leftmargin=*}
   \title{Born's Rule, EPR, and the Free Will Theorem}
   \author{Simon Kochen}
   
\begin{document}
   
   \maketitle
 
    \newcommand\cH{\mathcal{H}}
   \newcommand\cL{\mathcal{L}}
   \newcommand\rfl{\rotatebox{90}{$\rfloor$} }
   \newcommand\vp{\phi}
   \newcommand\tr{\mathop{\rm tr}\nolimits}
   \newcommand\Exp{\mathop{\rm Exp}\nolimits}
      \newcommand\Aut{\mathop{\rm Aut}\nolimits} 
   \newcommand\ssn[1]{\subsubsection*{#1}}
   \newcommand\R{\mathbb{R}}
   \newcommand\Ga{\Gamma}
   \newcommand\ga{\gamma}
      \newcommand\Q{\mathbb{Q}}
  \newcommand \lra[1]{\left\langle#1\right\rangle}
   \newcommand\la{\lambda}
   \newcommand\sig{\sigma}
   \newcommand\Om{\Omega}
   \newcommand\om{\omega}
    \newcommand\bpm{\begin{pmatrix}}
    \newcommand\epm{\end{pmatrix}}
   \newcommand\eps{\epsilon}
    \newcommand\bA{\mathbf{A}}
    \newcommand\bB{\mathbf{B}}
    \newcommand\ppsi{\psi} 
  \theoremstyle{definition}
\newtheorem*{defin}{Definition}
 
\section*{Introduction}

It is a historic irony that while Einstein devised the EPR experiment in order to show via special relativity that quantum mechanics is incomplete, many current physicists think the EPR experiment exhibits instantaneous effects that contradict strict Lorentz invariance. To be sure, these effects cannot be used to send superluminal signals, so most physicists feel free to ignore this glaring contradiction to special relativity.

This attitude seems incongruous to me, for although it is certainly a consequence of relativity that superluminal signals cannot be sent, this is far from its only significance. Lorentz invariance describes fundamental global symmetries of the universe. As Emmy Noether first showed, such symmetries imply the existence of basic observables. For instance, Galilean invariance leads to the existence of observables such as energy and momentum, whereas Lorentz invariance implies that energy and momentum are no longer observables; rather it is the four-vector consisting of a combination of energy and momentum that is now an observable.

Moreover, if the EPR experiment exhibits superluminal effects, then causality is violated. For in the Born form of EPR, if the two particles are space-like separated, and the measurement of the spin of the first particle has the instantaneous effect of giving a value to the spin of the second, then in a Lorentz frame in which the first is in the future of the second, we have backwards causation. Causality is not a consequence of the formalism in either classical or quantum physics, but is so imbedded in our thinking that it is generally accepted as a universal principle. In fact, causality has real consequences in physics, such as the Kramers-Kronig relations [1].

Not all physicists are willing to hide this apparent contradiction between quantum theory and relativity under the rug. Roger Penrose writes in his popular book ([2] p.446):
``The process does not seem to make sense at all when described in ordinary space-time terms. Consider a quantum state for a pair of particles. Such a state would normally be a correlated state. … Then an observation on one of the particles will affect the other in a non-local way which cannot be described in ordinary space-time terms consistently with special relativity (EPR: the Einstein-Podolsky-Rosen effect)."

I agree with Penrose that if the EPR experiment yields a superluminal effect, then it becomes necessary to re-think space-time geometry. However, I shall argue below that a careful application of  Born's Rule to EPR shows no such instantaneous effects, and that EPR is consistent with full Lorentz invariance. I then discuss how it is possible to understand the EPR correlations for non-commuting spin components in different directions without recourse to non-local effects. Finally, I use these results to discuss the Free Will Theorem. 

The arguments below are not the only ones against non-local effects of EPR. For instance, the paper [3] gives an argument based on a quantum erasure experiment to show the effects are local. Given the continuing controversy over non-locality, I may perhaps be forgiven for the careful exposition of well-known elementary topics.  

\section*{Born's Rule}
It will suffice to state the rule for a discrete observable $A$ of a system which is in a pure state given by the vector $\Omega$. Let $A = \sum \lambda_i P_i$  be the spectral decomposition of $A$. Then Born's rule says that the probability that a measurement of $A$ will give the value $\lambda_k$  is $<\Omega, P_k \Omega>$, and the system after the measurement is in the (unnormalized) state $P_k\Omega$ by the projection postulate.

In particular, if $\Omega$  is an eigenstate of $A$, so that $A \Omega = \lambda_k \Omega$ , say, then the observable $A$ has probability 1 of giving the value $\lambda_k$  and being in the state $P_k\Omega$  after the measurement of $A$. Thus, the observable $A$ is certain to give the value $\lambda_k$ if it is measured. Nothing is said about the value of $A$ before the measurement, and in particular that $A$ already has the value $\lambda_k$ \footnote{The same point is made in various text books. For instance, in [4] we see: ``Thus we can conclude that an eigenstate $\psi_j$ of an operator $P$ is a state in which the system yields with certainty the value $p_j$ [the eigenvalue belonging to $\psi_j$] \textit{when the observable corresponding to $P$ is measured}."       [italics added]}. If an eclipse is certain to happen tomorrow, we do not say that it has already happened.

Of course, if the state $\Omega$ has been prepared by a prior measurement of $A$, then $A$ already has the value $\lambda_k$, and the second measurement of $A$ would merely confirm that value. More generally, a prior measurement of an observable $A^\prime = \sum \lambda^\prime_i P_i$ with distinct $\lambda^\prime_i$, suffices. What is important is not the eigenvalues but the set of spectral projections $P_i$ , and the Boolean $\sigma$-algebra $B$ they generate.        

It is this Boolean algebra $B$, \textit{the interaction algebra}, that acquires truth values as a result of the measurement, namely, if the measurement gives the value $\lambda_k$, then $P_k$ is true and the other $P_i$'s are false, and this determines the truth values of the generated elements of $B$. 

That a measurement gives rise to a Boolean algebra of properties with truth values is not simply a consequence of quantum theory, but is, as noted by Kolmogorov in his classic book on probability [5], the case for any theory. Consider a classical experiment of throwing a die. There are six possible outcomes, the elementary events. Compound events, such as ``an even number of dots" or ``not two dots", are then among the 64 $(= 2^6)$ elements of the Boolean algebra generated by the elementary events. Similarly, the quantum Stern-Gerlach experiment to measure the $z$ component of spin of a spin 1 particle has three possible outcomes on the detector screen, which generate the eight element Boolean algebra of compound events. The outcomes correspond to properties of the measured system, in this case the spin properties $S_z = 0,+1,-1,$ given by the spectral decomposition of $S_z$.  These three properties generate a Boolean algebra, the interaction algebra, which is isomorphic to the Boolean algebra of events.  The actual spot on the detector determines the truth values of the compound events, and hence of all the elements of the interaction algebra of properties. In the case of an infinite number of properties the algebra generated forms a Boolean $\sigma$-algebra.(See [6] for details).

\section*{EPR}
We will discuss the EPR experiment in the Bohm form of two spin $\frac{1}{2}$ particles in the singlet state $\Gamma$ of total spin 0. Suppose that in that state the two particles are separated and the spin component $s_z$ of particle 1 is measured in some direction $z$. That means that the observable $s_z\otimes I$ of the combined system is being measured. 

 Let $P_z^\pm  = \frac{1}{2} I \pm s_z.$ We have the spectral decomposition $s_z\otimes I = (\frac{1}{2})P^+_z\otimes I + (-\frac{1}{2})P_z^-\otimes I,$                                                                                                                                                                                            so the interaction algebra is the Boolean algebra $B_1 = \{ P_z^+\otimes I, P_z^-\otimes I,1,0\}.$                                                                                                                                                                                                                               We may write the singlet state $\Gamma = \sqrt{\frac{1}{2}}(\psi^+_z\otimes \psi^-_z - \psi_z^- \otimes \psi_z^+),$ where $P_z^\pm \psi_z^\pm = \psi_z^\pm.$

Then  \\   $P_z^+\otimes I(\Gamma) = \sqrt{\frac{1}{2}} (\psi_z^+\otimes \psi_z^-)\\                                                                                                                                                                                                                                                                                                                        P_z^-\otimes I(\Gamma) = \sqrt{\frac{1}{2}} (\psi_z^- \otimes \psi_z^+).$

By Born's Rule, a measurement of $s_z\otimes I,$ the $z$-component of spin of particle 1, will yield the a value of $\frac{1}{2}$ with probability $\frac{1}{2}$ and a new state $\psi_z^+\otimes \psi_z^-,$ or a value of  $-\frac{1}{2}$ with probability $\frac{1}{2}$  and new state  $\psi_z^-\otimes \psi_z^+.$

                                                                                                                                                                                                  Now,  $\psi_z^+\otimes \psi_z^-$ and $\psi_z^-\otimes \psi_z^+$ are eigenstates of $I\otimes s_z$ with respective eigenvalues  $-\frac{1}{2}$ and $\frac{1}{2}$. It follows again from Born's Rule that if the $z$-spin component of particle 2 is measured, then it is certain to have the opposite $z$-component of spin.     
                                                                                                                                                                                       
The operator $I\otimes s_z$  has the spectral decomposition, $I\otimes s_z =  (\frac{1}{2})I\otimes  P_z^+ + (-\frac{1}{2})I\otimes P_z^-,$ so the interaction algebra of a measurement of particle 2 is the Boolean algebra $B_2 = \{ I\otimes  P_z^+, I\otimes P_z^-, 1, 0\}.$\footnote{ The interaction algebra of a measurement of both $s_z\otimes I$ and $I\otimes s_z$ (in any temporal order) is the Boolean algebra $B_1\oplus  B_2,$ the 16 element  Boolean algebra generated by $P_z^+\otimes I,$  and  $I\otimes  P_z^-.$                                                                                                                                                                                                                                                                                                         }

Our conclusion is that if particle 1 is measured and its spin is, say, $\frac{1}{2}$ then if particle 2 is measured, it is certain to have the value $-\frac{1}{2}$. It does not mean that particle 2 has acquired the opposite spin before it is measured, as a result of the measurement on particle 1. Those who claim this are reverting to the classical notion of intrinsic properties. The interaction algebra $B_1$ does not contain the properties $I\otimes P_z^+$ or $I\otimes P_z^-$ . The spin components do not have values unless and until the appropriate interaction happens.                                                                                                                                            
    
As we see, Born's Rule gives us the correct temporal order of events. \emph{A measurement of particle 1 giving a value, say, of $+\frac{1}{2}$ does not imply that particle 2 has acquired a spin value  of $-\frac{1}{2}$. What it  implies is that if and when it is measured, particle 2 will be certain to yield a spin value of $-\frac{1}{2}$.   }

 We give two further arguments against non-locality. The first argument is by continuity. If after particle 1 is measured to have spin $\frac{1}{2}$ in the $z$ direction particle 2 is measured in some other direction, at an angle $\theta$ to the $z$-axis, it is easily seen that it will give a value $-\frac{1}{2}$ with probability $\cos^2{(\frac{\theta}{2})}$. It cannot possibly mean that the second particle already has spin in that direction since there is a non-zero probability of $\sin^2{(\frac{\theta}{2})}$ that the spin is $+\frac{1}{2}$. Now, if particle 2's spin does not have a value as $\theta$ tends to 0, it should, by continuity, not have a value at $\theta =  0$, in the $z$ direction.

Here is a second, related argument. A Hermitean operator decomposes  Hilbert space into a direct sum of its eigenspaces. The operator $s_z\otimes s_z$ decomposes the four dimensional Hilbert space of the two particles as a direct sum $V_1\oplus V_2$ of the two 2 dimensional eigenspaces with the eigenvalues $\frac{1}{2}$ and $-\frac{1}{2}$. These subspaces have respective bases  $\{\psi_z^+\otimes \psi_z^+, \psi_z^-\otimes \psi_z^-\}$ and $\{\psi_z^-\otimes\psi_z^+, \psi_z^+\otimes \psi_z^-\}$. Now consider the direct sum operator $s_d\oplus s_z$ , where $s_d$  is the spin operator in an arbitrary direction $d$. Note that  $I\otimes s_z =  s_z\oplus s_z$, and that $s_d\oplus s_z$ has eigenstates  $\psi_z^-\otimes \psi_z^+$ and $\psi_z^+\otimes\psi_z^-$ with eigenvalues $+\frac{1}{2}$ and $-\frac{1}{2}$. As in EPR, with the two particles in the singleton state $\Gamma$, measure $s_z\otimes I$. Then the pair of particles will enter an eigenstate of $s_d\oplus s_z$ with opposite eigenvalue to $s_z\otimes I.$  Born's Rule tells us that a measurement of $s_d\oplus s_z$ is certain to give this opposite value to $s_z\otimes I.$ If however one maintains that $I\otimes s_z$ acquires an opposite value even before its measurement, it seems equally reasonable that for all $d$ each of the $s_d\oplus s_z$ acquires such an opposite value. However, the operators $s_d\oplus s_z$ do not pairwise commute for different values of $d$ and so do not form a Boolean algebra, as we expect of events that have happened.

A final point: Born's Rule tells us that a given state assigns probabilities to the interaction algebra of every observable. Is it possible that Born's Rule does not exhaust the meaning of a state as a complex vector? After all, the rule gives the probability value $<\Gamma,  P \Gamma > = || P\Gamma ||^2$ to the new state. Could the complex vector contain more information than its absolute value squared, which possibly instantaneously effects the other particle? The answer is no, because of the following theorem. An assignment of a probability $p(P)$ to every projection which is a probability measure on every Boolean $\sigma$-algebra of projections determines a unique density operator w such that $p(P) = tr(wP)$. For pure states, which are extreme points of the convex set of such probabilities, $w$ is a rank 1 projection $P = | \Gamma ><\Gamma |,$ and $p(P) = < \Gamma, P \Gamma  >.$  This is a consequence of Gleason’s Theorem. (See [6] for details).  So an assignment of probabilities to the eigenvalues of all observables, as is done by Born's Rule, is sufficient to determine the complex vector, up to a phase factor, defining the state.                                                                                                      

\section*{How EPR correlations are possible in different directions}
An EPR correlation in a single direction $z$ is no more difficult to understand than a correlation of two classical spinning particles of total angular momentum zero. For the fixed direction $z$ the EPR correlation may exist simply because the spin components in the $z$ direction have opposite values, and they were set up to be so correlated by preparing the total spin component $S_z$ of both particles to be zero. 

Thus, if we prepare the two particles to be in the state given by $S =1$ and $ S_z = 0,$ i.e. $\Gamma = \sqrt{\frac{1}{2}}(\psi_z^+\otimes \psi_z^- + \psi_z^-\otimes \psi_z^+)$, then the above calculation of the EPR correlation applies here to show the same correlation of opposite spin $z$-components. No one can claim for this case that we have superluminal effects, for they only exhibit the correlations which were set up by the initial state that $S_z = 0$ in a single direction $z$.

The new feature of EPR that is so puzzling is that the singleton state $S = 0$ yields spin correlations in different directions, say $z$ and $x$, where the spins $s_z$ and $s_x$ in those directions do not commute, and so cannot both have values simultaneously. How can correlations between spin components of two particles subsist when these spin components do not have values?

To understand how this can happen it is necessary to distinguish between events that have already happened and future contingent events. Thus, for instance, if $a \vee b$  \footnote{$a^\prime$ means ``not $a$", $a\ .\ b$ means ``$a$ and $b$", $a \vee b$ means ``$a$ or $b$", and $a\leftrightarrow b$ means ``$a$ if and only if $b$".} is currently true, then either $a$ is true or $b$ is true. When future events are considered, this no longer the case: if $a \vee b$ is certain to happen, it is not the case that $a$ is certain to happen or $b$ is certain to take place. A simple example of this is a single spin $\frac{1}{2}$ particle. For the spin projections $P_z^\pm$ given by $s_z= \pm\frac{1}{2},$ we have $P_z^+\vee P_z^- = 1,$ so $s_z = \frac{1}{2} \vee s_z = -\frac{1}{2}$ is always certain, for every direction $z$, but neither $s_z = \frac{1}{2}$ nor $s_z = -\frac{1}{2}$ is certain.

Born's Rule deals with precisely this distinction. We have seen that a given state gives the probabilities of the interaction Boolean algebra of projections defined by any observable which may be measured. If a particular observable is measured, the interaction algebra, which gives the events that have actually happened, now takes truth values, and the system enters a projected new state, with the corresponding new probabilities for contingent future properties. 

This feature is not restricted to quantum mechanics, but is common to phenomena with uncertain outcomes. Events $a$ and $b$ may each have probabilities other than 1 of occurring, while $a \vee b$ has probability 1. A famous example is Aristotle's sea battle. In [7],Chapter 9, he wrote  
``A sea-fight must either take place to-morrow or not, but it is not necessary that it should take place to-morrow, neither is it necessary that it should not take place, yet it is necessary that it either should or should not take place tomorrow. Since propositions correspond with facts, it is evident that when in future events there is a real alternative, and a potentiality in contrary directions, the corresponding affirmation and denial have the same character."

For future contingencies Aristotle retains the Law of Excluded Middle, $a \vee a^\prime$, but drops the Principle of Bivalence, that a proposition is either true or false when it comes to necessary future truth. So for Aristotle the disjunction  $a\vee b$ may be certain and so have a truth value, although neither $a$ nor $b$, being uncertain, have a truth value. This extends to other logical connectives (which are definable from $\vee$ and $^\prime$). Thus, logical equivalence $a\leftrightarrow b$ (which may be defined as $(a\ .\ b) \vee (a^\prime \ . \ b^\prime))$ may be certain, while $a$ and $b$ have no truth value. For example, the correlation, given by the equivalence ``My cat stays indoors $\leftrightarrow$ it rains in Princeton" has a truth value (e.g. it may be false, if she stayed out in the rain last Tuesday) even though ``My cat stays indoors" and ``it rains in Princeton" are only assigned probabilities, depending on the weather and my cat's temperament.

Now let us consider the correlations of spin components in the $z$ and $x$ directions in the EPR experiment. These can be written as 
\begin{equation}
s_z\otimes I = \frac{1}{2} \leftrightarrow I\otimes s_x = -\frac{1}{2} \hspace{1em} (\mbox{ i.e. } P_z^+\otimes I \leftrightarrow I\otimes P_z^-)
\end{equation}
\begin{equation}
s_x\otimes I = \frac{1}{2} \leftrightarrow I\otimes s_x = -\frac{1}{2} \hspace{1em}(\mbox{ i.e. } P_z^+ \otimes I \leftrightarrow I \otimes P_x^-)
\end{equation}
                                                             
 Now, the Boolean functions (1) $P_z^+\otimes I\leftrightarrow I\otimes P_z^-$  and (2) $P_x^+\otimes I \leftrightarrow I\otimes P_x^-$ are projections, and a calculation shows that these projections commute. One way to measure (1) is to measure $s_z\otimes I$ and $I\otimes s_z$ and check that they have opposite spins. However, this precludes one checking (2) by measuring $s_x\otimes I$ and $I\otimes s_x$, since $s_z\otimes I$ and $s_x\otimes I$ do not commute. A calculation shows that projection (1) is identical to the projection $1-S_z^2$   (i.e.the property $S_z = 0$) and (2) is the projection $1- S_x^2$ (i.e. $S_x = 0$). Here $S_z$ and $S_x$ refer to the combined components of spin of the two particles in the $z$ and $x$ directions. Since $S_z^2$ and $S_x^2$ commute we can measure both and check whether (1) and (2) are simultaneously true. In fact, a calculation shows that the conjunction $S_z = 0\ .  \ S_x = 0$ is equal to the projection $S = 0.$ So a measurement of the total spin $S$ suffices to check whether (1) and (2) are simultaneously certain. In particular, if we prepare the two particles to have $S=0$, and separate them carefully so that no outside torque is applied, then the combined system will continue to have $S=0$, by the conservation of angular momentum. In this way we could check that both correlations (1) and (2) are true.  Of course, we would not determine the values of the spin components of the individual particles  in either the $z$ or $x$ directions this way, but that is the whole point of this discussion: that a compound statement such as a correlation, can have a truth value without the component statements having a value. Just as $a \vee b$ may be certain even though $a$ and $b$ are not, correlations (1) and (2) may be certain even when $s_z\otimes I$ and $s_x\otimes I$ are not.

 This way of understanding the EPR correlations for non-commutative spins is not an ad hoc approach to this particular experiment. In [6]  we show that any correlation between two systems can be understood in a like manner.                                

\section*{Discussion}
This is not simply a controversy about semantics, so that in the final analysis, there is no real difference between non-local effects and correlations. We have already mentioned that an eminent relativist, Penrose, was willing to countenance non-causal effects and a change in space-time geometry because of his belief in the non-local effects of EPR. Spin correlations are defined in spin space, and have nothing to do with spatial separation or non-locality. If two classical particles have total spin zero, then they will have opposite components of spin in any direction, whether they are separated or not. They have opposite spins because they were correlated by the condition of total spin zero. Two quantum spin particles of total spin 0 will equally have opposite spins when measured in any direction because they were set up to be correlated by having total spin zero, just as with classical correlations. The new quantum aspect is rooted in correlations such as (1) and (2) above, which are true even though the spins in those directions cannot both have truth values; but as we have argued above, any theory of future contingencies must allow for the truth of compound properties such as  $a\leftrightarrow b$ without their constituents having truth values.

How did the idea that EPR exhibits non-locality arise? I believe it is based on a conflation of the two concepts of state and property. If a spin $\frac{1}{2}$ particle is in the state $\psi_z^+$, we say that it is in a state of ``spin $\frac{1}{2}$ in the $z$ direction." This an innocent conflation of the state $\psi_z^+$ and the property $P_z^+.$   The reason that it is harmless  is that $s_z$ is a rank 1 operator, so that $P_z^+(= |\psi_z^+ >< \psi_z^+|)$ has a one-dimensional image, viz. the ray containing $\psi_z^+$, and preparing the state $\psi_z^+$ also affirms the truth of the property $P_z^+$ in the interaction algebra $\{ P_z^+, P_z^-, 1, 0\}$.

When we consider a pair of particles in EPR, the situation is different. The observable $s_z\otimes I$ is a rank 2 operator, and $P_z^+\otimes I$  has a 2-dimensional image. The resulting state $\psi_z^+\otimes\psi_z^-$ is a vector lying in this image plane. It is tempting to similarly conflate this state with the conjunction of the properties $s_z\otimes I  =  +\frac{1}{2}$ and $I\otimes s_z = -\frac{1}{2}$, i.e. a state of spin $+\frac{1}{2}$  for particle 1 and spin $-\frac{1}{2}$ for particle 2. However, only the first particle has been measured, to give truth values to the interaction algebra $\{ P_z^+\otimes I, P_z^-\otimes I, 1, 0\}.$ All we can say of particle 2 is that it is certain to have the property $I\otimes  P_z^-$ of spin $-\frac{1}{2}$ if and when it is measured.  

\section*{EPR and the Free Will Theorem}
In [8] we proved the Free Will Theorem (FWT) on the basis of three axioms. One of these, the TWIN Axiom, was the EPR experiment adapted to the spin 1 case. We used Lorentz invariance via the third axiom, MIN, to give a contradiction to the assumption that the particles’ responses are determined by the past. It is clearly important to the proof that, as we have argued above, there are no superluminal effects in the application of the TWIN axiom. We shall now give the relevant part of the proof of the FWT that uses the EPR axiom TWIN.

At the same time, we recast the FWT by replacing MIN by a new axiom, LIN (Lorentz Invariance). Axiom LIN has the advantage over MIN in separating out the free will of the experimenters, and in dealing with a single experimenter rather than both. LIN is experimentally verifiable in principle, and is, I believe, universally accepted as true.

The axiom LIN says that the result of an experiment is Lorentz covariant: a change of Lorentz frames does not change the results of the experiment. This principle is called Lorentz Symmetry, and has been tested to a high degree of accuracy. (See [9].) To see what this means in the particular setting of the FWT, consider a Stern-Gerlach measurement of a spin $\frac{1}{2}$ particle. To see the covariance in a simple geometric way, we think of the detector screen in the form of diamond, i.e. a square rotated 45 degrees, with a horizontal diagonal that separates the spin up from the spin down spots on the detector. Now a change of Lorentz frame will change the square to a parallelogram, but the diagonal will remain a diagonal, and the spin up and down spots will remain above and below the diagonal in the new frame. It is in this sense that the result of the Stern-Gerlach experiment is Lorentz covariant. A similar covariance for measurement of squared spin components in the axiom SPIN is the content of LIN.

We now state our new axioms.

{\bf SPIN Axiom:} Measurements of the squared components of spin of a spin 1 particle in three orthogonal directions give the answers 1, 0, 1 in some order.

{\bf TWIN Axiom:} For twinned spin 1 particles of total spin $0$, suppose experimenter $A$ performs a triple experiment of measuring the squared spin components of particle $a$ in three orthogonal directions $x, y, z$, while experimenter $B$ measures the squared spin component of particle $b$ in one direction $w$. Then if $w$ happens to be in the same direction as one of $x, y, z,$ experimenter $B$'s measurement will necessarily yield the same answer as the corresponding measurement of $A$.

{\bf LIN Axiom:} For each of the 40 orthogonal triple experiments of $A$, particle $a$'s response as recorded on the detector screen does not depend on the inertial frame with respect to which the response is recorded. Similarly, for each of the 33 single experiments of $B$, particle $b$'s response on the detector is independent of the inertial frame.

(The 33 experiments of $B$ refer to the measurement of $S_z^2$ in the 33 directions $z$ specified in [8], and the 40 experiments of $A$ refer to the triple experiments measuring $(S_x^2, S_y^2, S_z^2)$ in the 40 orthogonal frames $(x,y,z)$ formed out of the 33 directions.)

{\bf The Free Will Theorem:}  The axioms SPIN, TWIN and LIN imply that if the experimenters $A$ and $B$ are free to choose from the respective 40 and 33 experiments, the response of each of the spin 1 particles is also free, in the sense that the response is not a function of properties outside its future light cone, i.e. of that part of the universe that is earlier than this response with respect to any given inertial frame.

{\bf Proof:}  We divide the proof into two parts: in the first part, the reduction of domain of the two functions $\Theta_a^F$ and $\Theta_b^G$ defined below, we copy the proof in [8], which did not use MIN. In the second part we use LIN in place of MIN to show that these two functions do not exist.

(1) We suppose to the contrary that particle $a$'s response $(i, j, k)$ to the triple experiment with directions $x, y, z$ is given by a function of properties $\alpha, \ldots$ that are earlier than this response with respect to some inertial frame $F$. We write this as \[\Theta_a^F(\alpha) = \mbox{ one of } (0,1,1),(1,0,1),(1,1,0)\] (in which only a typical one of the properties $\alpha$ is indicated). Similarly we suppose that $b$'s response $0$ or $1$ for the direction $w$ is given by a function \[\Theta^G_b(\beta)= \mbox{ one of } 0 \mbox{ or } 1\] of properties $\beta, \ldots$ that are earlier with respect to a possibly different inertial frame $G$  \footnote{Both in the EPR experiment and here $a$'s response may well be a function of future events, such as $b$'s response. In EPR, $b$'s future response of, say, $ -\frac{1}{2}$  implies that $a$'s present response is $+\frac{1}{2}$. Thus $a$'s response is conditioned on $b$'s future response. The conditional probability $p(x|y)$ of $x$ given $y$ is by its definition as $p(x\ .\ y)/p(y)$ independent of the time order of $x$ and $y$, and, in particular, so is the case when this probability is 1, so that $y$ implies $x$. We are here interested not in conditionality, but rather in not violating causality, and so we rule out future events by fiat in the definition of $\theta_a^F(\alpha)$ and $\theta_b^G(\beta).$  }.

(i) If either one of these functions, say $\Theta_a^F,$ is influenced by some information that is free in the above sense (i.e. not a function of $A$'s choice of directions and events $F$-earlier than that choice), then there must be an earliest (``infimum") $F$-time $t_0$ after which all such information is available to $a$. Since the non-free information is also available at $t_0$, all these information bits, free and non-free, must have a value $0$ or $1$ to enter as arguments in the function $\Theta_a^F.$ So we regard $a$'s response as having started at $t_0$. 

If indeed, there is any free bit that influences $a$, the universe has by definition taken a free decision near $a$ by time $t_0$, and we remove the pedantry by ascribing this decision to particle $a$.

(ii) From now on we can suppose that no such new information bits influence the particles' responses, and therefore that $\alpha$ and $\beta$ are functions of the respective experimenters' choices and of events earlier than those choices. 

Now an $\alpha$ can be expected to vary with $x, y, z$ and may or may not vary with $w$. However, whether the function varies with them or not, we can introduce all of $x, y, z, w$ as new arguments and rewrite $\Theta_a^F$ as a new function (which for convenience we give the same name) 
\begin{equation}\tag{*}
\Theta_a^F(x, y, z, w; \alpha^\prime)
\end{equation}
of $x,y,z,w$ and properties $\alpha^\prime$ independent of $x,y,z,w.$

To see this, replace any $\alpha$ that does depend on $x,y,z,w$ by the constant values $\alpha_1, \ldots, \alpha_{1320}$ it takes for the $40\times 33 = 1320$ particular quadruples $x,y,z,w$ we shall use. Alternatively, if each $\alpha$ is some function $\alpha(x,y,z,w)$ of $x,y,z,w,$ we may substitute these functions in (*) to obtain information bits independent of $x,y,z,w.$

Similarly, we can rewrite $\Theta_b^G$ as a function \[\Theta_b^G(x,y,z,w; \beta^\prime)\] of $x,y,z,w$ and properties $\beta^\prime$ independent of $x,y,z,w$. 

Now for the particle choice of $w$ that $B$ will make, there is a value $\beta_0$ for $\beta^\prime$ for which \[\Theta_b^G(x,y,z,w; \beta_0)\] is defined. By the above independence of $\beta^\prime$ from $x,y,z,w$, the function $\Theta_b^G(x,y,z,w; \beta_0)$ is defined with the same value $\beta_0$ for all $33$ values of $w$, so we may write it as \[\Theta^G_b(x,y,z,w).\]

Similarly there is a value $\alpha_0$ of $\alpha^\prime$ for which the function \[\Theta^F_a(x,y,z,w) = \Theta^F_a(x,y,z,w;\alpha_0)\] is defined for all 40 triples $x,y,z$ and it is also independent of $\alpha_0$, which argument we have therefore omitted.

(2) Now assume that $A$ and $B$ are space-like separated.  For an inertial frame $H$ in which $B$'s choice of the direction $w$ and $b$'s response occurs later than $A$'s choice of $x,y,z$ and $a$'s response, it follows from the definition of $\Theta_a^H$ as a function of properties that are earlier than $a$'s response in the frame $H$ that $w$ is not in the domain of $\Theta_a^H.$ Hence, by LIN, the function $\Theta_a^F$ is also independent of $w$, and we may write it as $\Theta_a^F(x,y,z).$ By similar reasoning, the function $\Theta_b^G$ is independent of $x,y,z$ and we may write it as $\Theta_b^G(w).$

By TWIN,                                              \[\theta_a^F(x,y,z)  =  (\theta_b^G(x), \theta_b^G(y), \theta_b^G(z)).\]                                                                                                                                                              

 Since $\theta_a^F(x,y,z)$ takes the value $(1,0,1), (1,1,0),$ or $(0,1,1),$ it follows that $\theta_b^G$ is a function of the given 33 directions which takes the values 1,0,1 in some order on every one of the 40 orthogonal triples $(x,y,z).$ This contradicts the Kochen-Specker Theorem as proved in [8], and shows that our assumption that the response of the particles is determined is false, and proves the theorem.

\end{document}